\documentclass[final]{ias2}
\usepackage{graphicx} 
\usepackage{multirow}
\usepackage{array} 

\usepackage{hyperref} 

\begin{document}

\markboth{Charged Anisotropic Star on Paraboloidal Spacetime}{B. S. Ratanpal, Jaita Sharma}

\title{Charged Anisotropic Star on Paraboloidal Spacetime}
\author[sin,ain]{B. S. Ratanpal} 
\email{bharatratanpal@gmail.com}
\author[sin, ain]{Jaita Sharma}
\email{jaita.sharma@gmail.com}
\address[sin,ain]{Department of Applied Mathematics, Faculty of Technology \& Engineering, The M. S. University of Baroda, Vadodara - 390 001, Gujarat, India}

\begin{abstract}
The charged anisotropic star on paraboloidal spacetime is reported by choosing particular form of radial pressure and electric field intensity. The non-singular solution of Einstein-Maxwell system of equation have been derived and it is shown that model satisfy all the physical plausibility conditions. It is observed that in the absence of electric field intensity, model reduces to particular case of uncharged Sharma \& Ratanpal model.
It is also observed that the parameter used in electric field intensity directly effects the mass of the star.
\end{abstract}



%
%

\maketitle




\section{Introduction}	
Theoretical investigation of Ruderman\cite{Ruder} leads to the conclusion that matter may be anisotropic in densities of order $10^{15}\;gm\;cm^{-3}$. The impact of anisotropy on steller configuration may be found in pioneering works of Bower and Liang\cite{BL} and Herrera and Santos\cite{HS}. Anisotropy may occur due to existance of type 3A superfluids\cite {Ruder,BL,KW}, or phase transition\cite{Soko}. Consenza \textit{et. al.}\cite{CHEW} developed the procedure to obtain anisotropic solutions from isotropic solutions of Einstein's field equations. Tikekar and Thomas\cite{TT} found exact solutions of Einstein's field equations for anisotropic fluid sphere on pseudo spheroidal spacetime. The key feature of their model is the high variation of density from centre to boundary of stellar configuration. The class of exact anisotropic solutions on spherically symmetric spacetime has been obtained by Mak and Harko\cite{MH}. Karmakar \textit{et. al.}\cite{KMSM} analysed the role of pressure anisotropy for Vaidya-Tikekar\cite{VT} model. Paul \textit{et. al}\cite{PCKT} developed anisotropic stellar model for strage star. A core-envelope model describing superdense stars with anisotropic fluid distribution has been obtained by Thomas and Ratanpal\cite{TR}, Thomas \textit{et. al.}\cite{TRV}  \& Tikekar and Thomas\cite{TT1}. Hence the study of anisotropic fluid distribution is important in general theory of relativity.\\\\
\noindent The study of Einstein-Maxwell system is carried out by several authors. Patel and Kopper\cite{PK} obtained charged analog of Vaidya-Tikekar\cite{VT} solution. The study of analytic models of quark stars have been carrried out by Komathiraj and Maharaj\cite{KM}, they found a class of solutions of Einstein-Maxwell system. Charged anisotropic matter with linear equation of state have been extensively studied by Thirukkanesh and Maharaj\cite{TM}.\\\\
\noindent Hence, both anisotropy and electromagnatic field are important in relativistic astrophysics. In this paper charged anisotropic model of a stellar configuration have been studied on the background of paraboloidal spacetime. Sec-2 describes, the field equations for charged static stallar configuration on paraboloidal spacetime and their solution is obtained. Sec-3 described the physical plausibility condition and Sec-4 contains discussion.

\section{Field Equations and Solution}
The interior of stellar configuration is described by static spherically symmetric paraboloidal spacetime metric,
\begin{equation}\label{ISpaceTime1}
	ds^{2}=e^{\nu(r)}dt^{2}-\left(1+\frac{r^{2}}{R^{2}} \right)dr^{2}-r^{2}\left(d\theta^{2}+\sin^{2}\theta d\phi^{2}\right),
\end{equation} 
with the energy-momentum tensor for anisotropic charged fluid,
\begin{equation}\label{EMTensor}
	T_{ij}=diag\left(\rho+E^{2},\;p_{r}-E^{2},\;p_{t}+E^{2},\;p_{t}+E^{2} \right),
\end{equation}
where $\rho$ is the energy density, $p_{r}$ is the radial pressure, $p_{t}$ is the tangential pressure and $E$ is the electric field intensity. These quantities are measured relative to the comoving fluid velocity $u^{i}=e^{-\nu}\delta_{0}^{i}$. For the spacetime metric (\ref{ISpaceTime1}) and energy-momentum tensor (\ref{EMTensor}), the Einstein-Maxwell system takes the form,
\begin{equation}\label{rho1}
	\rho+E^{2}=\frac{3+\frac{r^{2}}{R^{2}}}{R^{2}\left(1+\frac{r^{2}}{R^{2}} \right)^{2}},
\end{equation}
\begin{equation}\label{pr1}
	p_{r}-E^{2}=\frac{\nu'}{r\left(1+\frac{r^{2}}{R^{2}}\right)}-\frac{1}{R^{2}\left(1+\frac{r^{2}}{R^{2}}\right)},
\end{equation}
\begin{equation}\label{pt1}
	p_{t}+E^{2}=\frac{1}{1+\frac{r^{2}}{R^{2}}}\left[\frac{\nu''}{2}+\frac{\nu'^{2}}{4}+\frac{\nu'}{2r}\right]-\frac{\nu' r}{2R^{2}\left(1+\frac{r^{2}}{R^{2}}\right)^{2}}-\frac{1}{R^{2}\left(1+\frac{r^{2}}{R^{2}}\right)^{2}},
\end{equation}
\begin{equation}\label{Sigma1}
	\sigma=\frac{\left(r^{2}E\right)'}{r^{2}\sqrt{1+\frac{r^{2}}{R^{2}}}},
\end{equation}
where $\sigma$ is the proper charge density and prime denotes differentiation with respect to $r$. In field equations (\ref{rho1}) - (\ref{Sigma1}), velocity of light $c$ is taken as $1$ also $\frac{8\pi G}{c^{4}}=1$.\\\\
The anisotropic parameter $\Delta$ is defined as, 
\begin{equation}\label{Delta1}
	\Delta=p_{t}-p_{r}.
\end{equation}
To solve the system (\ref{rho1}) - (\ref{Sigma1}) radial pressure is assumed to be of the form,
\begin{equation}\label{pr2}
	p_{r}=\frac{p_{0}\left(1-\frac{r^{2}}{R^{2}}\right)}{R^{2}\left(1+\frac{r^{2}}{R^{2}}\right)^{2}},
\end{equation}
where $p_{0}>0$ is the model parameter and $\frac{p_{0}}{R^{2}}$ is central pressure. At the boundary of the star $r=R$, $p_{r}$ must vanish, which gives $r=R$ as the radius of the star.\\\\
This form of radial pressure is prescribed by Sharma and Ratanpal\cite{SR} to describe anisotropic stellar model admitting a quadratic equation of state on paraboloidal spacetime. Equations (\ref{pr2}) and (\ref{pr1}) gives,
\begin{equation}\label{NuDash1}
	\nu'=\frac{p_{0}r\left(1-\frac{r^{2}}{R^{2}}\right)}{R^{2}\left(1+\frac{r^{2}}{R^{2}}\right)}+\frac{r}{R^{2}}-E^{2}r\left(1+\frac{r^{2}}{R^{2}}\right).
\end{equation}
We assume electric field intensity of the form,
\begin{equation}\label{E1}
	E^{2}=\frac{k\frac{r^{2}}{R^{2}}}{R^{2}\left(1+\frac{r^{2}}{R^{2}} \right)^{2}},
\end{equation}
where $k\geq 0$ is a model parameter, from equation (\ref{E1}) it is clear that $E$ is decreasing in radially outward direction. Equations (\ref{NuDash1}) and (\ref{E1}) leads to,
\begin{equation}\label{NuDash2}
	\nu'=\frac{\left(2p_{0}+k\right)r}{R^{2}\left(1+\frac{r^{2}}{R^{2}}\right)}+\left(1-p_{0}-k\right)\frac{r}{R^{2}},
\end{equation}
and hence,
\begin{equation}\label{Nu}
	\nu=\log\left[C\left(1+\frac{r^{2}}{R^{2}}\right)^{\left(\frac{2p_{0}+k}{2}\right)}\right]+\left(\frac{1-p_{0}-k}{2}\right)\frac{r^{2}}{R^{2}},
\end{equation}
where $C$ is constant of integration. Therefore spacetime metric (\ref{ISpaceTime1}) is written as,
\begin{equation}\label{ISpaceTime2}
	ds^{2}=C\left(1+\frac{r^{2}}{R^{2}}\right)^{\left(\frac{2p_{0}+k}{2} \right)}e^{\left(\frac{1-p_{0}-k}{2}\right)\frac{r^{2}}{R^{2}}}dt^{2}-\left(1+\frac{r^{2}}{R^{2}} \right)dr^{2}-r^{2}\left(d\theta^{2}+\sin^{2}\theta d\phi^{2} \right).
\end{equation}
The spacetime metric (\ref{ISpaceTime2}) should continously match with Reissner-Nordstrom spacetime metric
\begin{equation}\label{ESpaceTime}
	ds^{2}=\left(1-\frac{2m}{r}+\frac{Q^{2}}{r^{2}}\right)dt^{2}-\left(1-\frac{2m}{r}+\frac{Q^{2}}{r^{2}}\right)^{-1}dr^{2}-r^{2}\left(d\theta^{2}+\sin^{2}\theta\right)d\phi^{2},
\end{equation}
at the boundary of the star $r=R$, where $\left(p_{r}\right)\left(r=R\right)=0$. This matching conditions gives
\begin{equation}\label{M}
	M=\frac{k+2R^{2}}{8R},
\end{equation}
and
\begin{equation}\label{C}
	C=\frac{e^{\left(\frac{p_{0}+k-1}{2}\right)}}{\left(2p_{0}+2+k\right)/2},
\end{equation}
where $M$ is the mass enclosed the spherical body of radius $R$, hence the electric field intensity parameter $k$ directly effects the mass of the star.
Equations (\ref{pr1}), (\ref{pt1}), (\ref{Delta1}) and (\ref{NuDash2}) gives anisotropic parameter $\Delta$ as,
\begin{equation}\label{Delta2}
	\Delta=\frac{\frac{r^{2}}{R^{2}}\left[X_{1}+Y_{1}\frac{r^{2}}{R^{2}}+Z_{1}\frac{r^{4}}{R^{4}}\right]}{4R^{2}\left(1+\frac{r^{2}}{R^{2}}\right)^{3}},
\end{equation}
which vanishes at $r=0$, where $X_{1}=p_{0}^{2}-8p_{0}-12k+3$, $Y_{1}=-2p_{0}^{2}-2p_{0}k+2p_{0}-8k+4$, $Z_{1}=1+p_{0}^{2}+k^{2}-2p_{0}-2k+2p_{0}k$. Equations (\ref{rho1}) and (\ref{E1}) gives,
\begin{equation}\label{rho2}
	\rho=\frac{3+\left(1-k\right)\frac{r^{2}}{R^{2}}}{R^{2}\left(1+\frac{r^{2}}{R^{2}}\right)^{2}},
\end{equation}
and from equations (\ref{Delta1}), (\ref{pr2}) and (\ref{Delta2}) we get expression of $p_{t}$ as,
\begin{equation}\label{pt2}
	p_{t}=\frac{4p_{0}+X_{1}\frac{r^{2}}{R^{2}}+Y_{1}\frac{r^{4}}{R^{4}}+Z_{1}\frac{r^{6}}{R^{6}}}{4R^{2}\left(1+\frac{r^{2}}{R^{2}}\right)^{3}}.
\end{equation}
Hence equations (\ref{rho2}), (\ref{pr2}), (\ref{pt2}), (\ref{E1}) and (\ref{Delta2}) describes matter density, radial pressure, tangential pressure, electric field intensity and measure of anisotropy respectively.
\section{Physical Plausibility Conditions}
Following Delgaty and Lake\cite{Delgaty} we impose the following condition on the sytem to make model physically acceptable.
\begin{itemize}
\item [(i)] $\rho(r),~p_{r}(r),~p_{t}(r) \geq 0 $ for $ 0 \leq r \leq R$.
\item [(ii)] $\rho-p_{r}-2p_{t} \geq 0$ for $ 0 \leq r \leq R$.
\item [(iii)] $\frac{d\rho}{dr},~\frac{dp_{r}}{dr},~\frac{dp_{t}}{dr} < 0$ for $0 \leq r \leq R$.
\item [(iv)] $0 \leq \frac{dp_{r}}{d\rho} \leq 1$; $0 \leq \frac{dp_{t}}{d\rho} \leq 1$, for $0 \leq r \leq R$.
\end{itemize}
From equation (\ref{rho2}), $\rho(r=0)=\frac{3}{R^{2}}>0$ and $\rho(r=R)=\frac{4-k}{4R^{2}}$,
therefore, $\rho>0$ for $0\leq r\leq R$ if $k\leq4$, i.e.
\begin{equation}\label{k1}
	0\leq k\leq 4.
\end{equation}
From equation (\ref{pr2}) $p_{r}(r=0)=\frac{p_{0}}{R^{2}}>0$ as $p_{0}>0$ and $p_{r}(r=R)=0$. Hence $p_{r}\geq 0$ for $0\leq r\leq R$. It is required that $p_{t}\geq 0$ for $0\leq r\leq R$ and further to get the simple bounds on $p_{0}$ and $k$, we assume $p_{r}=p_{t}$ at $r=R$.\\\\
from (\ref{pt2}), $p_{t}(r=0)=\frac{p_{0}}{R^{2}}>0$ as $p_{0}>0$, and $p_{t}(r=R)=\frac{k^{2}-22k+8-8p_{0}}{32R^{2}}$, at $r=R$, $p_{t}=p_{r}=0$ if,
\begin{equation}\label{p0}
	p_{0}=\frac{k^{2}-22k+8}{8},
\end{equation}
but $k$ should be chosen such that $p_{0}$ is positive, which restrict the value of $k$ as,
\begin{equation}\label{k2}
	k<0.3699.
\end{equation}
Hence,
\begin{equation}\label{k3}
	0\leq k<0.3699,\;\;\;\;\;p_{0}=\frac{k^{2}-22k+8}{8}
\end{equation}
which is condition for positivity of $p_{t}$.\\\\
Hence condition (i) is satisfied throughout the star. For the values of $k$ and $p_{0}$ specified in (\ref{k3}), programatically it has been varified that condition (ii) i.e. energy condition is satisfied throughout the star. From equation (\ref{rho2}),
\begin{equation}\label{drhodr}
	\frac{d\rho}{dr}=-\frac{2r}{R^{4}}\frac{\left[(5+k)+(1-k)\frac{r^{2}}{R^{2}}\right]}{\left(1+\frac{r^{2}}{R^{2}} \right)^{3}},
\end{equation}
from equation (\ref{drhodr}), $\left(\frac{d\rho}{dr}\right)(r=0)=0$ and $\left(\frac{d\rho}{dr}\right)(r=R)=-\frac{3}{2R^{3}}<0$. Hence $\rho$ is decreasing throught the star. From equation (\ref{pr2}),
\begin{equation}\label{dprdr}
	\frac{dp_{r}}{dr}=\frac{-2p_{0}r\left(3-\frac{r^{2}}{R^{2}}\right)}{R^{4}\left(1+\frac{r^{2}}{R^{2}}\right)^{3}}.
\end{equation}
Now, $\left(\frac{dp_{r}}{dr}\right)(r=0)=0$ and $\left(\frac{dp_{r}}{dr}\right)(r=R)=\frac{-p_{0}}{2R^{3}}<0$ as $p_{0}>0$. Hence $p_{r}$ is decreasing throughout the star. From equation (\ref{pt2}),
\begin{equation}\label{dptdr}
	\frac{dp_{t}}{dr}=\frac{r\left[X_{2}+Y_{2}\frac{r^{2}}{R^{2}}+Z_{2}\frac{r^{4}}{R^{4}}\right]}{2R^{4}\left(1+\frac{r^{2}}{R^{2}}\right)^{4}},
\end{equation}
where $X_{2}=p_{0}^{2}-20p_{0}-12k+3$, $Y_{2}=-6p_{0}^{2}+12p_{0}-4p_{0}k+8k+2$ and $Z_{2}=5p_{0}^{2}-4p_{0}+8p_{0}k+3k^{2}+2k-1$. Now $\left(\frac{dp_{t}}{dr}\right)(r=0)=0$ and $\left(\frac{dp_{t}}{dr}\right)(r=R)=\frac{-12p_{0}+3k^{2}-2k+4p_{0}k+4}{32R^{3}}$. Substituting $p_{0}$ from equation (\ref{p0}) in $\left(\frac{dp_{t}}{dr}\right)(r=R)$, $\left(\frac{dp_{t}}{dr}\right)(r=R)<0$ if $4k^{3}-76k^{2}+280k-64<0$. This further restrict value of $k$ as $k<0.2446$.\\\\
Hence, if
\begin{equation}\label{k4}
	0\leq k<0.2446,\;\;\;\;\;p_{0}=\frac{k^{2}-22k+8}{8},
\end{equation}
then $\frac{d\rho}{dr}$, $\frac{dp_{r}}{dr}$ and $\frac{dp_{t}}{dr}$ are decreasing in radially outward direction between $0\leq r\leq R$. From equations (\ref{drhodr}) and (\ref{dprdr}) we have,
\begin{equation}\label{dprdrho}
	\frac{dp_{r}}{d\rho}=\frac{p_{0}\left(3-\frac{r^{2}}{R^{2}}\right)}{\left[(5+k)+(1-k)\frac{r^{2}}{R^{2}}\right]}.
\end{equation}
At the centre of the star $\left(\frac{dp_{r}}{d\rho}\right)(r=0)<1$ if $k<24.8810$, which is consistent with condition (\ref{k4}) and at the boundary of the star $\left(\frac{dp_{r}}{d\rho}\right)(r=R)<1$ if $k<22.7047$, which is also consistent with conditon (\ref{k4}). From equations (\ref{drhodr}) and (\ref{dptdr}) we have,
\begin{equation}\label{dptdrho}
	\frac{dp_{t}}{d\rho}=\frac{-\left[X_{2}+Y_{2}\frac{r^{2}}{R^{2}}+Z_{2}\frac{r^{4}}{R^{4}}\right]}{4\left(1+\frac{r^{2}}{R^{2}}\right)\left[(5+k)+(1-k)\frac{r^{2}}{R^{2}}\right]}.
\end{equation}
Now, $\left(\frac{dp_{t}}{d\rho}\right)(r=0)<1$ if $k<19.4283$, which is consistent with condition (\ref{k4}) and $\left(\frac{dp_{t}}{d\rho}\right)(r=R)<1$ if $k<6.6371$, which is also consistent with condition (\ref{k4}). Hence for $0\leq k<0.2446$ and $p_{0}=\frac{k^{2}-22k+8}{8}$, all the physical plausibility conditions are satisfied.
\section{Dicussion}
Certain ascepts of charged relativistic star on paraboloidal spacetime have been discussed. It is observed that all the physical plausibility conditions are satisfied for $0\leq k\leq 0.2446$ and $p_{0}=\frac{k^{2}-22k+8}{8}$. The plots of $\rho$ (charged, uncharged), $p_{r}$ \& $p_{t}$ (charged), $p_{r}$ \& $p_{t}$ (uncharged), anisotropy $\Delta$ (charged, uncharged), $\rho-p_{r}-2p_{t}$ (charged, uncharged), $\frac{dp_{r}}{d\rho}$ \& $\frac{dp_{t}}{d\rho}$ (Charged), $\frac{dp_{r}}{d\rho}$ \& $\frac{dp_{t}}{d\rho}$ (uncharged) over $\frac{r^{2}}{R^{2}}$ for $R=10$, $k=0.2$ and taking $G=c^{2}=1$ are shown in figures 1 to 7. It is observed that energy condition is satisfied throughout the star. When $k=0$ the value of $p_{0}=1$ and the model reduces to the Sharma \& Ratanpal\cite{SR} model. Hence the model described here is charged generalization of particular case $p_{0}=1$ of uncharged Sharma \& Ratanpal\cite{SR} model.
\begin{figure}[h]
\begin{center}
\includegraphics[width=12cm]{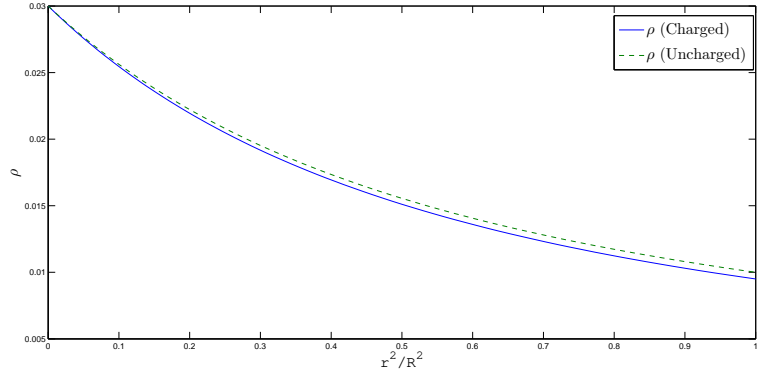}\\
\caption{Variation of density ($\rho$) (charged \& uncharged) against $\frac{r^{2}}{R^{2}}$.}\label{rho}
\end{center}
\end{figure}

\pagebreak
\begin{figure}[h]
\begin{center}
\includegraphics[width=12cm]{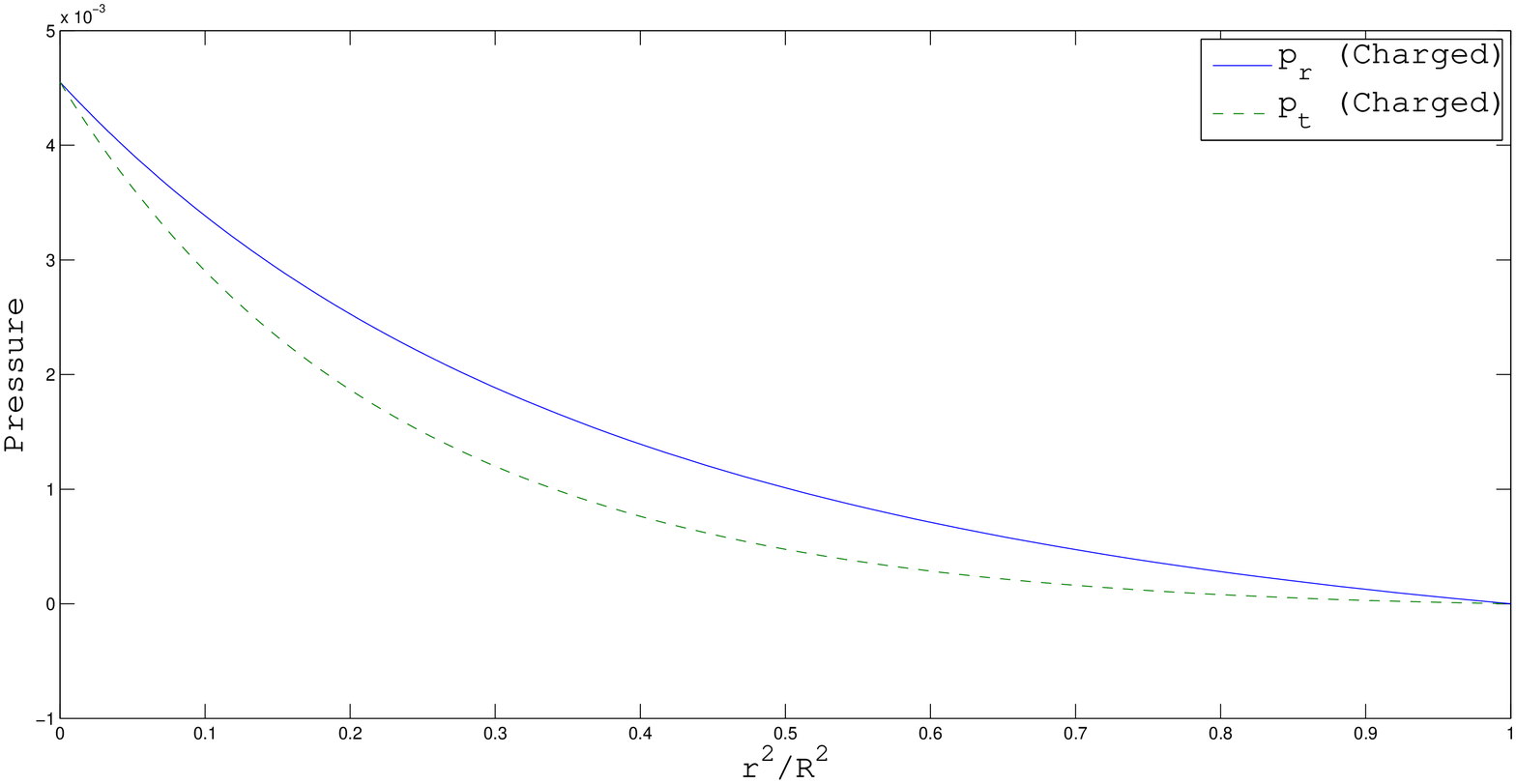}\\
\caption{Variation of radial pressure ($p_{r}$) \& tangential pressure ($p_{t}$) (charged) against $\frac{r^{2}}{R^{2}}$.}\label{pressureC}
\end{center}
\end{figure}

\begin{figure}[h]
\begin{center}
\includegraphics[width=12cm]{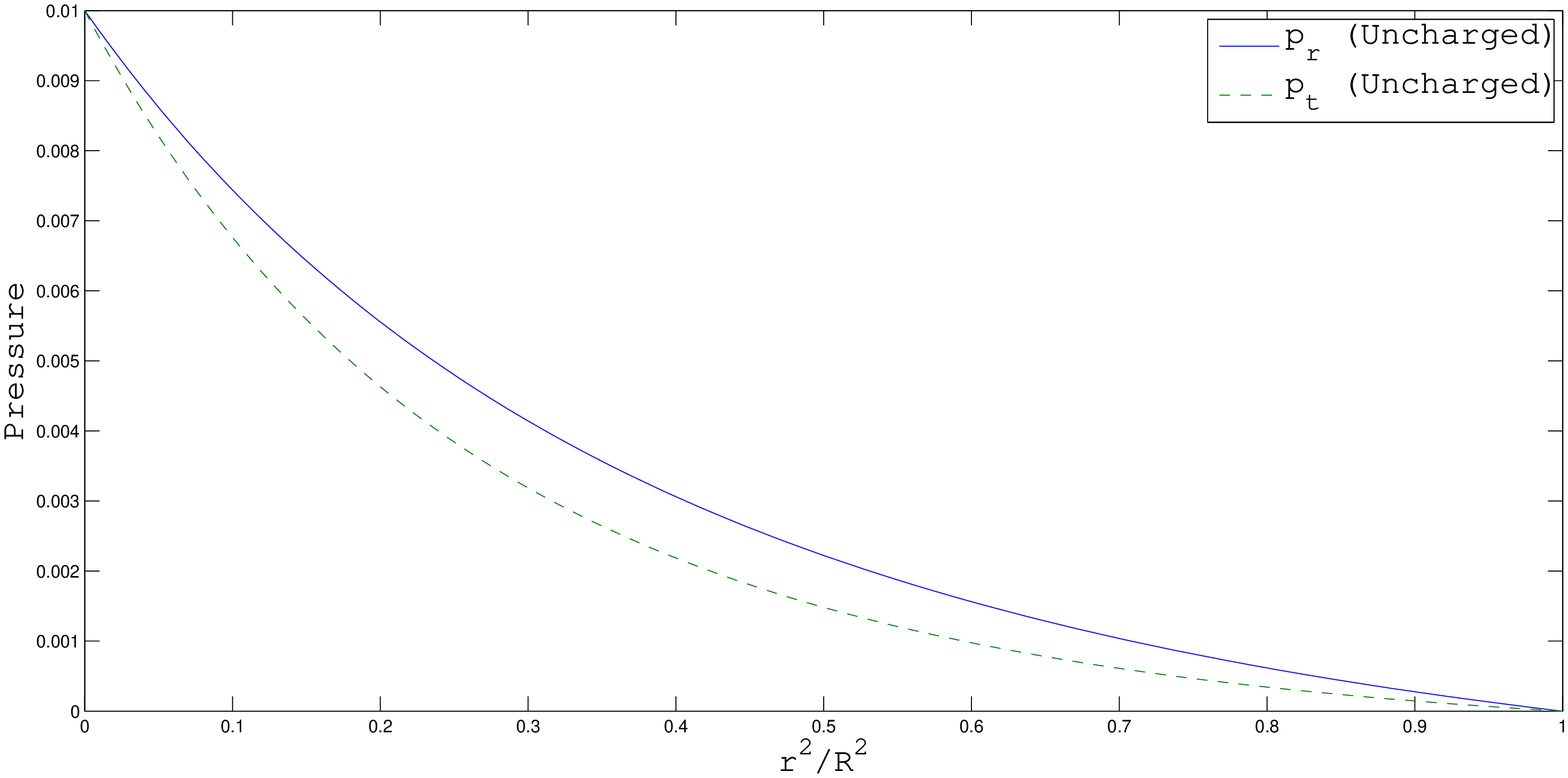}\\
\caption{Variation of radial pressure ($p_{r}$) \& tangential pressure ($p_{t}$) (uncharged) against $\frac{r^{2}}{R^{2}}$.}\label{pressureU}
\end{center}
\end{figure}

\pagebreak
\begin{figure}[h]
\begin{center}
\includegraphics[width=12cm]{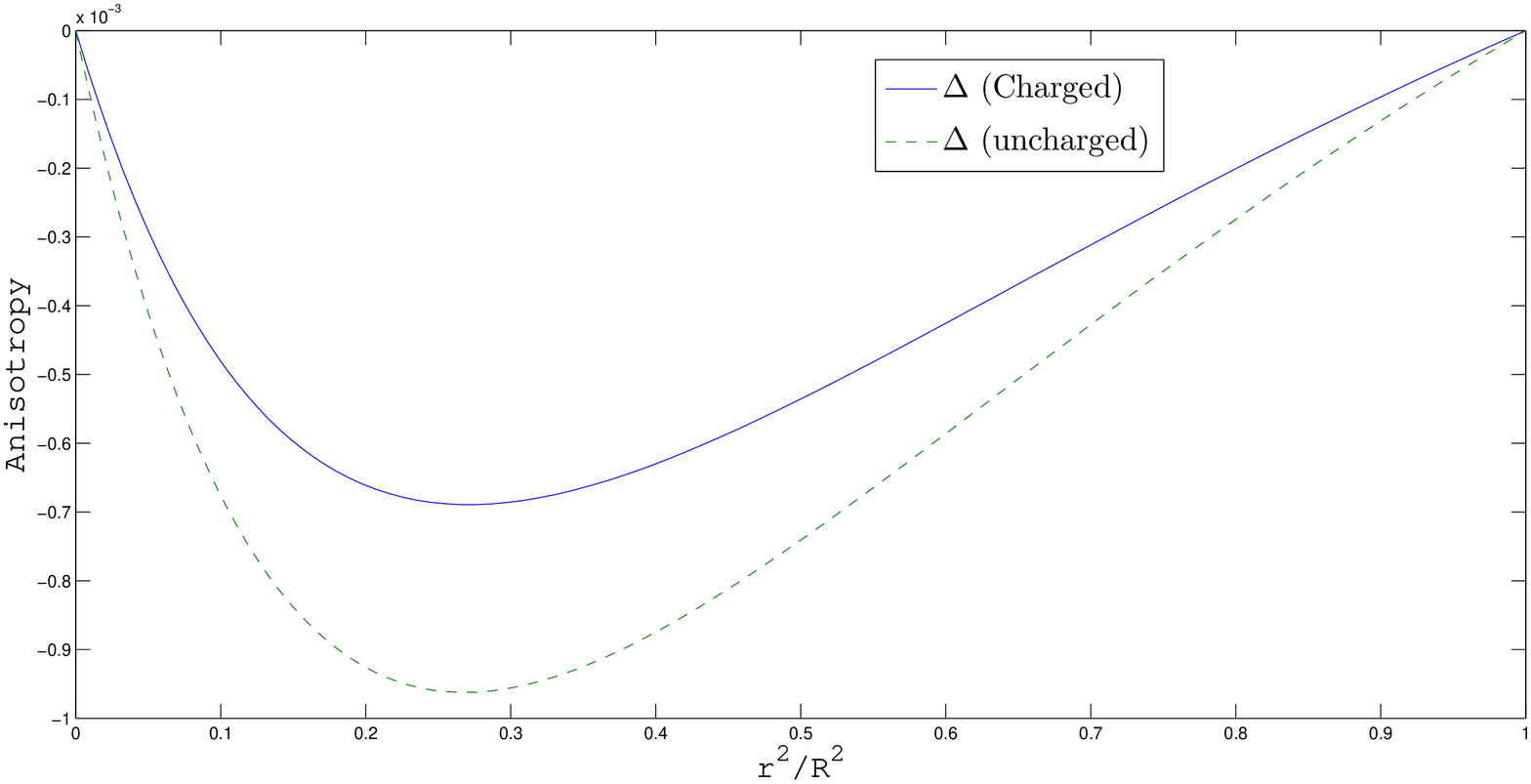}\\
\caption{Variation of anisotropy ($\Delta$) (charged \& uncharged) against $\frac{r^{2}}{R^{2}}$.}\label{anisotropy}
\end{center}
\end{figure}

\begin{figure}[h]
\begin{center}
\includegraphics[width=12cm]{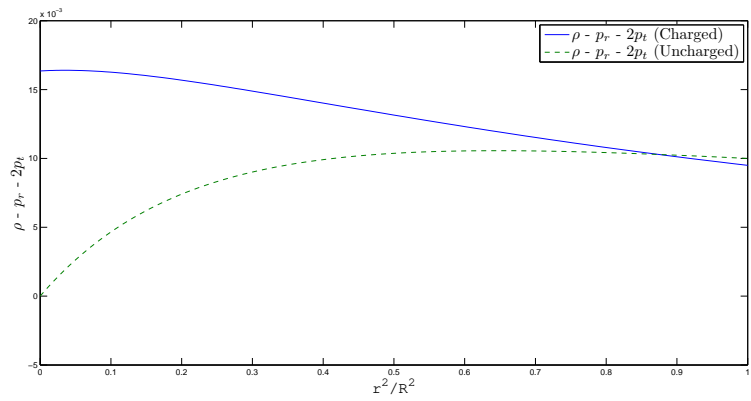}\\
\caption{Variation of energy condition ($\rho-p_{r}-2p_{t}$) (charged \& uncharged) against $\frac{r^{2}}{R^{2}}$.}\label{anisotropy}
\end{center}
\end{figure}

\pagebreak
\begin{figure}[h]
\begin{center}
\includegraphics[width=12cm]{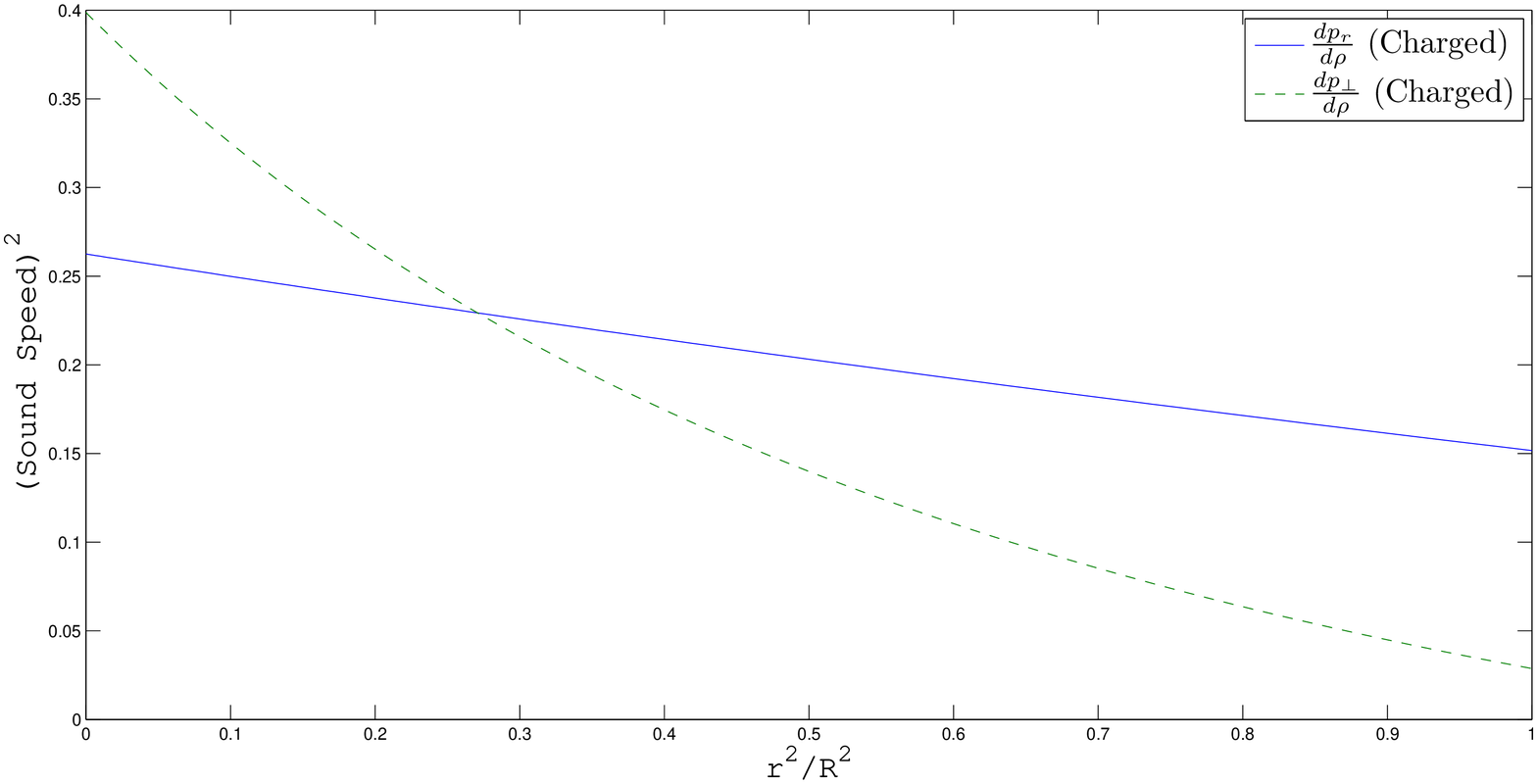}\\
\caption{Variation of $\frac{dp_{r}}{d\rho}$ \& $\frac{dp_{t}}{d\rho}$ (charged) against $\frac{r^{2}}{R^{2}}$.}\label{soundspeedC}
\end{center}
\end{figure}

\begin{figure}[h]
\begin{center}
\includegraphics[width=12cm]{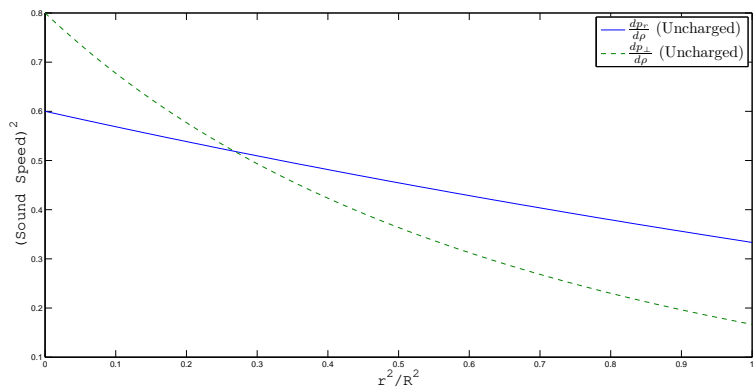}\\
\caption{Variation of $\frac{dp_{r}}{d\rho}$ \& $\frac{dp_{t}}{d\rho}$ (uncharged) against $\frac{r^{2}}{R^{2}}$.}\label{soundspeedC}
\end{center}
\end{figure}

\pagebreak


\begin{thebibliography}{00}  


\bibitem{Ruder} R. Ruderman, {\it Annu. Rev. Astron. Astrophys.} {\bf 10} (1972) 427.
\bibitem{BL} R L Bowers and E P T Liang, {\it Astrophysical J.} {\bf 188} (1974) 657.
\bibitem{HS} L Herrera and N O Santos, {\it Phys. Rep.} {\bf 286} (1997) 53.
\bibitem{KW} R Kippenhahn and A Weigert, Stellar Structure and Evolution (Springer Verlang, Berlin Heidelberg, New York) (1990).
\bibitem{Soko} A I Sokolov, {\it JETP} {\bf 52} (1980) 575.
\bibitem{CHEW} M Cosenza, L Herrera, M Esculpi and L Witten, {\it J. Math. Phys.} {\bf 22} (1981) 118.
\bibitem{TT} R Tikekar and V O Thomas, {\it Pramana J. Phys.} {\bf 52} (1999) 237.
\bibitem{MH} M K Mak and T Harko, {\it Proc. R. Soc. Lond. A} {\bf 459} (2003) 393.
\bibitem{KMSM} S Karmakar, S Mukherjee, R Sharma and S D Maharaj, {\it Pramana J. Phys.} {\bf 68} (2007) 881.
\bibitem{VT} P C Vaidya and R Tikekar, {\it J. Astrophys. Astr.} {\bf 3} (1982) 325.
\bibitem{PCKT} B C Paul, P K Chattopadhyay, S Karmakar and R Tikekar, {\it Mod. Phys. Lett. A} {\bf 26} (2011) 575.
\bibitem{TR} V O Thomas and B S Ratanpal, {\it Int. J. Mod. Phys. D} {\bf 16} (2007) 1479.
\bibitem{TRV} V O Thomas, B S Ratanpal and P C Vinodkumar, {\it Int. J. Mod. Phys. D} {\bf 14} (2005) 85.
\bibitem{TT1} R Tikekar and V O Thomas, {\it Pramana J. Phys.} {\bf 64} (2005) 5.
\bibitem{PK} L K Patel and S S Kopppar, {\it Ast. J. Phys.} {\bf 40} (1987) 441.
\bibitem{KM} K Komathiraj and S D Maharaj, {\it Int. J. Mod. Phys. D} {\bf 16} (2007) 1803.
\bibitem{TM} S Thirukkanesh and S D Maharaj, {\it Class. Quantum Grav.} {\bf 25} (2008) 235001-1.
\bibitem{SR}R Sharma and B S Ratanpal, {\it Int. J. Mod. Phys. D} {\bf 22} (2013) 1350074-1.
\bibitem{Delgaty}M S R Delgaty and K Lake, {\it Computer Physics Communications} {\bf 115} (1998) 395.

\end{thebibliography}
\end{document}